\documentstyle[prl,aps,floats,epsf,color]{revtex}
\begin{document}
\baselineskip=12pt
\def\be{\begin{equation}}
\def\ee{\end{equation}}
\def\bea{\begin{eqnarray}}
\def\eea{\end{eqnarray}}
\def\E{{\rm e}}
\def\bearst{\begin{eqnarray*}}
\def\eearst{\end{eqnarray*}}
\def\peleven{\parbox{11cm}}
\def\peffec{\peight{\bearst\eearst}\hfill\peleven}
\def\pspace{\peight{\bearst\eearst}\hfill}
\def\ptwelve{\parbox{12cm}}
\def\peight{\parbox{8mm}}
\twocolumn[\hsize\textwidth\columnwidth\hsize\csname@twocolumnfalse\endcsname

\title
{ Localization of elastic waves in heterogeneous media with
off-diagonal disorder and long-range correlations}

\author
{F. Shahbazi,$^1$ Alireza Bahraminasab,$^2$ S. Mehdi Vaez
Allaei,$^3$ Muhammad Sahimi,$^{4}$ and M. Reza Rahimi
Tabar$^{2,5}$ }

\vskip 1cm

\address
{\it $^1$Department of Physics, Isfahan University of Technology,
Isfahan 84156, Iran\\
$^2$Department of Physics, Sharif University of Technology, Tehran
11365-9161, Iran\\
$^3$Institute for Advanced Studies in Basic Sciences, Gava Zang,
Zanjan, 45195-159, Iran\\
$^4$Department of Chemical Engineering, University of Southern
California, Los Angeles, California 90089-1211\\
$^5$CNRS UMR 6529, Observatoire de la C$\hat o$te d'Azur, BP 4229,
06304 Nice Cedex 4, France}
 \maketitle


\begin{abstract}

Using the Martin-Siggia-Rose method, we study propagation of
acoustic waves in strongly heterogeneous media which are
characterized by a broad distribution of the elastic constants.
Gaussian-white distributed elastic constants, as well as those
with long-range correlations with non-decaying power-law
correlation functions, are considered. The study is motivated in
part by a recent discovery that the elastic moduli of rock at
large length scales may be characterized by long-range power-law
correlation functions. Depending on the disorder, the
renormalization group (RG) flows exhibit a transition to localized
regime in {\it any} dimension. We have numerically checked the RG
results using the transfer-matrix method and direct numerical
simulations for one- and two-dimensional systems, respectively.

\end{abstract}
\hspace{.3in}
\newpage
] Understanding how waves propagate in heterogeneous media is
fundamental to such important problems as earthquakes, underground
nuclear explosions, the morphology of oil and gas reservoirs,
oceanography, and medical and materials sciences [1]. For example,
seismic wave propagation and reflection are used to not only
estimate the hydrocarbon content of a potential oil or gas field,
but also to image structures located over a wide area, ranging
from the Earth's near surface to the deeper crust and upper
mantle. The same essential concepts and techniques are used in
such diverse fields as materials science and medicine.

In condensed matter physics, a related problem, namely, the nature
of electronic states in disordered materials, has been studied for
several decades and shown to depend strongly on the spatial
dimensionality $d$ of the materials [2]. It was rigorously shown
that, for one-dimensional (1D) systems, even infinitesimally small
disorder is sufficient for localizing the wave function,
irrespective of the energy [3], and that the envelop of the wave
function $\psi(r)$ decays exponentially at large distances $r$
from the domain's center, $\psi(r)\sim\exp(-r/\xi)$, with $\xi$
being the localization length. The most important results for
$d>1$ follow from the scaling theory of localization [4] which
predicts that, for $d\leq 2$, all electronic states are localized
for any degree of disorder, while a transition to extended states
- the metal-insulator transition - occurs for $d>2$ if disorder is
sufficiently strong. The transition between the two states is
characterized by divergence of the localization length,
$\xi\propto|W-W_c|^{-\nu}$, where $W_c$ is the critical value of
the disorder intensity. Wegner [5] derived a field-theoretic
formulation for the localization problem which, together with the
scaling theory [4], predict a lower critical dimension, $d_c=2$,
for the localization problem. These predictions have been
confirmed by numerical simulations [6].

Wave characteristics of electrons suggest that the localization
phenomenon may occur in other wave propagation processes. For
example, consider propagation of seismic waves in heterogeneous
rock. In this case, the interference of the waves that have
undergone multiple scattering, caused by the heterogeneities of
the medium, may cause their localization. Unlike electrons,
however, the classical waves do not interact with one another and,
therefore, propagation of such waves in heterogeneous media (such
as porous rock) provides [7] an ideal model for studying the
classical Anderson localization [8-12] in strongly disordered
media. This is the focus of this Letter. We study localization of
acoustic waves in strongly heterogeneous media, and formulate a
field-theoretic method to investigate the problem in the media
that are characterized by a broad distribution of the elastic
constants. Localization of acoustic waves was previously studied
by several groups [13], although not in the strongly disordered
media that we consider in this Letter. The system that we study is
the continuum limit of an acoustic system with off-diagonal
disorder. Our approach is based on the method first introduced by
Martin, Siggia and Rose [14] for analyzing dynamical critical
phenomena. We calculate the one-loop beta functions [8,14] for
both spatially delta-correlated and power-law correlated disorder
in the elastic constants, and show that in any case there is a
disorder-induced transition from delocalized to localized states
for {\it any} $d$. In addition to be interesting on its own, our
study is motivated in part by the recent discovery [15] that the
distribution of the elastic moduli of heterogeneous rock contains
long-range correlations characterized by a {\it nondecaying}
power-law correlation function. Baluni and Willemsen [16] studied
propagation of acoustic waves in a 1D layered system, which can be
thought of as a simple model of rock (although their goal was not
to study acoustic wave propagation in rock), and showed that the
waves are localized. However, they did not consider
higher-dimensional systems, nor did they study the type of
disorder that we consider in the present Letter. The possibility
of wave localization in disordered media with long-range,
nondecaying correlations has important practical implications
which we will discuss briefly. However, our results are completely
general and apply to any material in which the local elastic
constants are distributed broadly. We confirm the analytical
predictions for 1D and 2D systems using numerical simulations.

Wave propagation in a medium with a distribution of elastic
constants is described by the following equation (for simplicity
we consider the scalar wave equation):

\begin{equation}
\frac{\partial^2}{\partial t^2}\psi({\bf
x},t)-\mbox{\boldmath$\nabla$}\cdot \left[\lambda({\bf
x})\mbox{\boldmath$\nabla$}\psi({\bf x},t)\right]=0\;,
\end{equation}

where $\psi({\bf x},t)$ is the wave amplitude, and $\lambda({\bf
x})= C({\bf x})/m$ is the ratio of the elastic stiffness $C({\bf
x})$ and the mean density $m$ of the medium. We then write
$\lambda$ as,

\begin{equation}
\lambda({\bf x})=\lambda_0+\eta({\bf x})\;,
\end{equation}

where $\lambda_0=\langle\lambda({\bf x})\rangle$. In this Letter
we assume $\eta({\bf x})$ to be a Gaussian random process with a
zero mean and the covariance,

\begin{eqnarray}
\langle\eta({\bf x})\eta({\bf x}')\rangle&&=2K(|{\bf x}-{\bf
x}'|)=2D_0\delta^d ({\bf x}-{\bf x}')\cr \nonumber \\
&&+2D_\rho|{\bf x}-{\bf x}'|^{2\rho-d}.
\end{eqnarray}

in which $D_0$ and $D_\rho$ represent the strength of the disorder
due to the delta-correlated and power-law correlated parts of the
disorder. Previously, Souillard and co-workers [17] studied wave
propagation in disordered fractal media, which is characterized by
a {\it decaying} power-law correlation function. Their study is
not, however, directly related to our work. Consider a wave
component $\psi({\bf x},\omega)$ with angular frequency $\omega$,
which is obtained by taking the temporal Fourier transform of Eq.
(1) which yields the following equation for propagation of a wave
component in a disordered medium,

\begin{equation}
\nabla^2\psi({\bf x},\omega)+\frac{\omega^2}{\lambda_0}\psi({\bf
x},\omega)+ \mbox{\boldmath$\nabla$}\cdot\left[\frac{\eta({\bf
x})}{\lambda_0} \mbox{\boldmath$\nabla$}\psi({\bf
x},\omega)\right]=0\;.
\end{equation}

Since $\eta({\bf x})$ is a Gaussian variable, we obtain a
Martin-Siggia-Rose effective action $S_e$ for the probability
density functional of the wave function $\psi({\bf x},\omega)$,
given by
\begin{eqnarray}
&&S_e(\psi_I,\psi_R,\tilde\psi,\chi,\chi^*=\cr \nonumber
\\ && \int d{\bf x}d{\bf x}' [(i\tilde\psi_I({\bf
x}')(\nabla^2+\frac{\omega^2}{\lambda_0}) \psi_I({\bf x}) \cr
\nonumber \\ && + i \tilde \psi_R({\bf
x}')(\nabla^2+\frac{\omega^2} {\lambda_0})\psi_R({\bf
x}). \cr \nonumber \\
 &&+\chi^*({\bf
x}')(\nabla^2+\frac{\omega^2}{\lambda_0}) \chi({\bf
x}))\delta({\bf x}-{\bf x}') \cr \nonumber
\\ &&+(i\mbox{\boldmath$\nabla$}
\tilde\psi_I\mbox{\boldmath$\nabla$}\psi_I+i\mbox{\boldmath$\nabla$}
\tilde\psi_R\mbox{\boldmath$\nabla$}\psi_R+\mbox{\boldmath$\nabla$}\chi
\mbox{\boldmath$\nabla$}\chi)\frac{K({\bf x}-{\bf
x}')}{\lambda_0^2}\cr \nonumber \\
&&
\times\left(i\mbox{\boldmath$\nabla$}\tilde\psi_I\mbox{\boldmath$\nabla$}
\psi_I+i\mbox{\boldmath$\nabla$}\tilde\psi_R\mbox{\boldmath$\nabla$}\psi_R+
\mbox{\boldmath$\nabla$}\chi\mbox{\boldmath$\nabla$}\chi\right)]\;.
\end{eqnarray}
Here, $\tilde\psi_I({\bf x})$, $\tilde\psi_R({\bf x})$, $\chi$ and
$\chi^*$ are the auxiliary and Grassmanian fields of the
field-theoretic formulation, respectively. Two coupling constants,
$g_0=D_0/\lambda_0^2$, and, $g_\rho= D_\rho/\lambda_\rho^2$,
appear in $S_e$. Thus, we carry out a renormalization group (RG)
analysis in the critical limit, $\omega^2/\lambda_0\to 0$, to
derive, to one-loop order, the beta functions [7,12] that govern
the two couplings under the RG transformation. The results are
given by,

\begin{equation}
\beta(\tilde{g}_0)=\frac{\partial\tilde{g}_0}{\partial\ln
l}=-d\tilde{g}_0
+8\tilde{g}_0^2+10\tilde{g}_\rho^2+20\tilde{g}_0\tilde{g}_\rho\;,
\end{equation}
\begin{equation}
\beta(\tilde{g}_\rho)=\frac{\partial\tilde{g}_\rho}{\partial\ln
l}=(2\rho-d)\tilde{g}_\rho+12\tilde{g}_0\tilde{g}_\rho+16\tilde{g}_\rho\;,
\end{equation}

where $l>1$ is the re-scaling parameter, and $\tilde{g}_0$ and
$\tilde{g}_\rho$ are given by,

\begin{equation}
\tilde{g}_0=k_d\left[\frac{d+5}{2d(d+2)}\right]g_0\;,
\end{equation}

\begin{equation}
\tilde{g}_\rho=k_d\left[\frac{d+5}{2d(d+2)}\right]g_\rho\;,
\end{equation}
with $k_d=S_d/(2\pi^d)$, and $S_d$ being the surface area of the
$d-$dimensional unit sphere. Examining the RG flows, Eqs. (6) and
(7), reveals that, depending on $\rho$, there are two distinct
regimes:

(i) For $0<\rho<d/2$ there are three sets of fixed points: The
trivial Gaussian fixed point ($g^*_0=g_\rho^*=0$) which is stable,
and two non-trivial fixed points and eigendirections. One is,
$\{g_0^*=d/8,\;g_\rho^*=0\}$, while the other set is given by,

\begin{eqnarray}
g_0^*&&=-\frac{4}{41}\left[d+\frac{5}{16}(2\rho-d)\right] \cr
\nonumber
\\&&-\frac{4}{41}\sqrt{\left[d+\frac{5}{16}(2\rho-d)\right]^2+\frac{205}{256}
(2\rho-d)^2} \cr \nonumber
\\ && g_\rho^*=\frac{3}{4}g_0^*+\frac{1}{16}(d-2\rho)\;,
\end{eqnarray}

which is stable in one eigendirection but unstable in the other
eigendirection.

The corresponding RG flow diagram is shown in Figure 1. Therefore,
for $0<\rho<d/2$ the one-loop RG calculation indicates that the
system with uncorrelated disorder is unstable against long-range
correlated disorder toward a new fixed point in the coupling
constants space, for which there is a phase transition from
delocalized to localized states with increasing the disorder
intensity.

(ii) For $\rho>d/2$ there are two fixed points: the Gaussian fixed
point which is stable on the $g_0$ axis but unstable on the
$g_\rho$ axis, and the

non-trivial fixed point, $\{g^*_0=d/8,\;g^*_\rho=0\}$, which is
unstable in all directions. The RG flow diagram for this case is
shown in Figure 2. The implication is that, while the power-low
correlated disorder is relevant, no new fixed point exists to
one-loop order and, therefore, the long-wavelength behavior of the
system is determined by the long-range component of the disorder.
This means that for $\rho>d/2$ the waves are localized for {\it
any} $d$. In addition, in both cases (i) and (ii) the system
undergoes a disorder-induced transition when only the uncorrelated
disorder is present. Let us mention that the above results are
general so long as $D_\rho>0$ (which is the only physically
acceptable limit). For $D_\rho<0$ the above phase space is valid
for $\rho>\frac{1}{2}(d+1)$.

\begin{figure}
\epsfxsize=6truecm\epsfbox{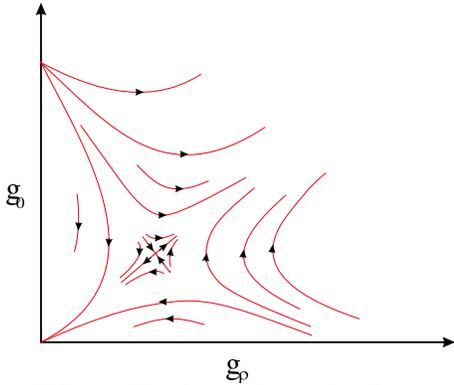}
 \narrowtext \caption{ The
figure shows the flows in the coupling constants space for
$0<\rho<d/2$. The below one is  for $\rho>d/2$.}
 \end{figure}

\begin{figure}
\epsfxsize=6truecm\epsfbox{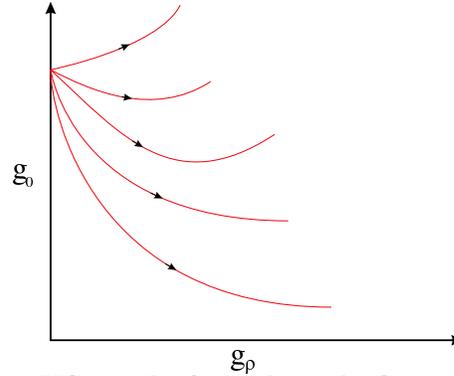}
 \narrowtext \caption{The
figure shows the flows in the coupling constants space for
$\rho>d/2$.}
 \end{figure}

\begin{figure}
\epsfxsize=6truecm\epsfbox{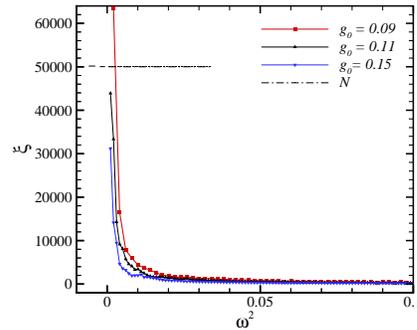}
 \narrowtext \caption{The localization length $\xi$ for a disordered chain with
50,000 site and its dependence on the frequency $\omega$}
 \end{figure}
\begin{figure}
\epsfxsize=6truecm\epsfbox{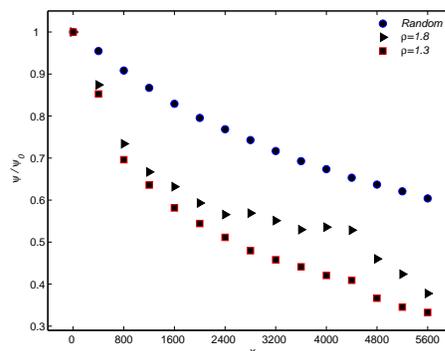}
 \narrowtext \caption{ Amplitude of the waves in 2D systems, for both randomly
and uniformly-distributed and power-law correlated elastic
constants. }
 \end{figure}
To test these predictions, we have carried out numerical
simulations of the problem in both 1D and 2D. Consider first the
1D disordered systems. In this case, the waves are localized when
the wave functions are of the form,
$\psi(x)=f(x)\exp(-|x-x_0|/\xi)$, where $f(x)$ is a stochastic
function which depends on the particular realization of the
disordered chain, and $\xi$ is the localization length.
Experimentally, the simplest 1D model that exhibits wave
localization is [18] a 15 m long steel wire with a 0.178 mm
diameter, suspended vertically. The tension in the wire is
maintained with a weight attached at its lower end. The function
$\psi(x,t)$ consists of transverse waves in the wire with an
electromechanical actuator at one end of the wire.

It was shown [18] that, even for very small deviations (less than
1\%) from periodicity, the diagonal disorder (e.g., variations in
the resonance frequencies of the oscillators) produces
localization (which is in agreement with Furstenberg's theorem
[19]), while variations (up to 13\%) in the sizes of the masses
(off-diagonal disorder) result in localization lengths that are
much larger than the size of the system. This is in agreement with
our theoretical prediction.

To reproduce this result numerically and to calculate the
localization length $\xi$, we used the transfer-matrix (TM) method
[12]. Discretizing Eq. (1) and writing down the result for site
$n$ of a linear chain yields,
\begin{equation}
(\omega+\lambda_n)\psi_n+\lambda_{n+1}\psi_{n+2}-(\lambda_{n+1}-\lambda_n)
\psi_{n+1}=0\;,
\end{equation}

which can be rewritten in the recursive form

\begin{equation}
M_n\left(\begin{array}{c}
\psi_{n+2} \\
\psi_{n+1}\\
\end{array}\right)=\left(\begin{array}{c}
\psi_{n+1}\\
\psi_{n}\\
\end{array}\right)
\end{equation}

with
\begin{equation}
M_n=\left(\begin{array}{cc} \displaystyle
-\frac{\omega^2-\lambda_n+\lambda_{n-2}}{\lambda_n} &
\displaystyle \frac{\lambda_{n-2}}{\lambda_n}\\
1 & 0 \\
\end{array}\right)\;.
\end{equation}

The localization length $\xi(\omega)$ is then defined by,
$\xi(\omega)^{-1}= \lim_{N\to\infty} N^{-1}|\psi_N/\psi_0|$, where
$N$ is the chain's length. For every realization of the disorder
we computed $\psi_N$ and, hence, $\xi(\omega)$. We chose,
$\psi_0=\psi_1=1/\sqrt{2}$, and averaged $\xi$ over a large
ensemble of realizations for a fixed system size $N$ and frequency
$\omega$. We then repeated this procedure for several values of
$N$ and $\omega$. The extended states correspond to having,
$\lim_{N\to\infty}\xi/N= {\rm constant}>1$. As $\xi$ is also a
function of $\omega$, we chose, $\omega= 2\pi\sqrt{\lambda_0}/N$,
which is the smallest mode of the system. Our RG analysis
indicates that this mode most likely passes through the 1D
disordered chain.

The TM computations indicate that the coupling constant $g(N)$
follows a finite-size scaling, $g(N)=g_0+1/N$, where $g_0$ is the
coupling constant in the thermodynamic ($N\to\infty$) limit; we
find that, $g_0\simeq 0.117$. In addition, when $\xi\to N$, one
can write, $N\propto (g-g_0)^{-\nu}$ and obtain an estimate of the
localization exponent $\nu$. Our analytical results indicated [20]
that in $d$ dimensions, $\nu=1/d$, which agrees with the TM
calculations that yield $\nu=1$. Figure 3 shows the results for
the localization length $\xi$ as a function of $\omega$. These
results confirm the RG predictions for disordered linear chains.

To further check the RG results, we also solved Eq. (1) in 2D
using the finite-difference method with second-order
discretization for both the space and time variables. Such
approximations are acceptable as we work in the limit of low
frequencies or long wavelengths. For short wavelengths we should
use higher-order discretizations for the spatial variables [20]. A
$L_x\times L_y$ grid was used with $L_x=8000$ and $L_y=400$. The
parameter $\lambda({\bf x})$, representing the local,
gridblock-scale elastic constant, was distributed with a power-law
correlation function, of the type considered above, with its
spatial distribution generated using the midpoint displacement
method [21]. We also simulated the case in which the local values
of $\lambda({\bf x})$ were uncorrelated and uniformly distributed,
with the same variance as that of the power-law case. By inserting
a wave source on a line on one side of the grid, we simulated
numerically 2D wave propagation through the system. Periodic
boundary conditions were imposed in the lateral direction, which
did not distort the nature of the wave propagation, as we used
large system sizes. The decay in the amplitude of the wave is
caused by scattering from heterogeneities of system generated by
the distribution of the local elastic constants. The accuracy of
the solution was checked by considering the stability criterion
and the wavelength of the source [22], and using higher-order
finite-difference discretizations. To compute the amplitude decay
in the medium, we collected the numerical results at 80 receivers
(grid points), distributed evenly throughout the grid, along the
direction of wave propagation. The results were averaged over 32
realizations of the system.

Figure 4 presents the decay in the wave amplitude through the
uniformly random medium, and that of a medium with a nondecaying
power-law correlation function for the local elastic constants
$\lambda({\bf x})$, with $\rho=1.3$ and 1.8. The wave amplitudes
for the correlated cases decline much faster than those in the
uniformly random medium. In particular, for $\rho=1.3$, which
corresponds to negative correlations (that is, a large local
elastic modulus is likely to be neighbor to a small one, and vice
versa), the amplitude decreases rather sharply. These results
confirm the RG predictions for 2D systems.

In summary, we show that, depending on the nature of disorder,
acoustic waves in strongly disordered media can be localized or
delocalized in {\it any} dimensions. In particular, they can be
extended in disordered 1D systems if the correlation function for
the distribution of the local elastic constants is of nondecaying
power-law type, and that the waves are localized in {\it any}
dimension if the exponent $\rho$ of the power-law correlation
function is larger than $d/2$. These results, which contradict the
generally-accepted view that off-diagonal disorder has a much
weaker effect on localization than the diagonal disorder, have
important practical implications. For example, in order for
seismic records to contain meaningful information on the geology
and content of a natural porous formation of linear size $L$, the
localization length $\xi$ must be larger $L$. Otherwise,
propagation and scattering of such waves can provide information
on the formation only up to length scale $\xi$; one cannot obtain
meaningful information at larger length scales [23]. The
localization length $\xi$ is, clearly, a function of the
dimensionality of the system, the exponent $\rho$ and amplitude
$D_\rho$, and other relevant physical parameters of the system.
Its determination remains a major task [20].

We thank John Cardy for useful comments. The work of SMVA was
supported by the NIOC.


\begin{thebibliography}{99}
\bibitem{20} The details will be given elsewhere.
\end{thebibliography}
\end{document}